\documentclass[twocolumn,nofootinbib,floatfix,10pt,aps]{revtex4-2}
\usepackage{amsmath}
\usepackage{amssymb}
\usepackage{wasysym}
\usepackage{graphicx}
\usepackage{color,soul}
\usepackage{physics}
\usepackage{siunitx}
\usepackage{dsfont}
\usepackage{float}
\usepackage{multirow}
\usepackage{mathdots} 
\usepackage[english]{babel}
\usepackage{blindtext}
\usepackage[english,nomargin,inline,marginclue,draft]{fixme}
\pdfpageheight\paperheight
\pdfpagewidth\paperwidth
\usepackage[colorlinks,linkcolor=blue,anchorcolor=blue,citecolor=blue,urlcolor=blue]{hyperref}
            
\fxusetheme{colorsig}
\FXRegisterAuthor{cg}{acg}{CG}  
\FXRegisterAuthor{th}{ath}{\color{blue}TH}  
\FXRegisterAuthor{ib}{aib}{\color{red}IB} 
\FXRegisterAuthor{sh}{ash}{\color{cyan}SH} 
\FXRegisterAuthor{db}{adb}{\color{green}DB} % now I can ..
\FXRegisterAuthor{ps}{aps}{PS}
\makeatletter
\renewcommand*\FXLayoutInline[3]{%
  {\@fxuseface{inline}\ignorespaces{\color{fx#1}[#3: #2]}}}
\makeatother

\long\def\symbolfootnote[#1]#2{\begingroup%
\def\thefootnote{\fnsymbol{footnote}}\footnotetext[#1]{#2}\endgroup}

\def\nobreakbefore{%
  \relax\ifvmode\else
    \ifhmode
      \ifdim\lastskip > 0pt\relax
        \unskip\nobreakspace
      \else % added to put a ~if no space was typed. (Unclear why it sometimes worked before )
        \nobreakspace
      \fi
    \fi
  \fi
}
\let\oldcite\cite
\renewcommand\cite{\nobreakbefore\oldcite}

\begin{document}
\title{A Rb-Cs dual-species magneto-optical trap}
% \captionsetup[figure]{name={FIG.},labelsep=period,justification=raggedright}

\author{Shi-Yao Shao$^{1,2}$}

\author{Qing Li$^{1,2}$}

\author{Li-Hua Zhang$^{1,2}$}

\author{Bang Liu$^{1,2}$}

\author{Zheng-Yuan Zhang$^{1,2}$}

\author{Qi-Feng Wang$^{1,2}$}

\author{Jun Zhang$^{1,2}$}

\author{Yu Ma$^{1,2}$}

\author{Tian-Yu Han$^{1,2}$}

\author{Han-Chao Chen$^{1,2}$}

\author{Jia-Dou Nan$^{1,2}$}

\author{Yi-Ming Yin$^{1,2}$}

\author{Dong-Yang Zhu$^{1,2}$}

\author{Ya-Jun Wang$^{1,2}$}

\author{Dong-Sheng Ding$^{1,2}$}

\email{dds@ustc.edu.cn}

\author{Bao-Sen Shi$^{1,2}$}
\affiliation{$^1$Key Laboratory of Quantum Information, University of Science and Technology of China, Hefei, Anhui 230026, China.}
\affiliation{$^2$Synergetic Innovation Center of Quantum Information and Quantum Physics, University of Science and Technology of China, Hefei, Anhui 230026, China.}
\affiliation{}
\date{\today}

\begin{abstract}
We describe a three-dimensional (3D) magneto-optical trap (MOT) capable of simultaneously capturing \textsuperscript{85}Rb and \textsuperscript{133}Cs atoms. Unlike conventional setups, our system utilizes two separate laser systems that are combined before entering the vacuum chamber, enabling the simultaneous trapping of two different atomic species. Additionally, in our 3D MOT configuration, two (of three) pairs of laser beams are not orthogonal to the chamber surfaces but are aligned at a 45° angle. With a total trapping laser power of 8 mW and repump laser power of 4 mW for Rb atoms, and a total trapping laser power of 7.5 mW and repump laser power of 1.5 mW for Cs atoms, we achieve optical depths (OD) of 3.71 for Rb and 3.45 for Cs, demonstrating efficient trapping for both species. Our 3D MOT setup allows full horizontal optical access to the trapped atomic ensembles without spatial interference from the trapping or repump laser beams. Moreover, the red detuning for trapping both atomic species is smaller than in traditional configurations. This system offers a versatile platform for exploring complex phenomena in ultracold atom physics, such as Rydberg molecule formation and interspecies interactions.

\end{abstract}

\maketitle

\section{Introduction}

The magneto-optical trap (MOT)\cite{raab1987trapping} is a key tool in the field of cold atom physics and the most common laser cooling method. Since its introduction in 1987\cite{raab1987trapping}, the MOT has been widely used in various fundamental and applied research areas. By combining laser cooling and magnetic trapping techniques, the MOT can cool atoms to microkelvin temperatures or even lower. These ultracold temperatures significantly reduce the thermal motion of atoms, enabling precise control and measurement of quantum states. As a result, the MOT has become an essential foundational tool in cutting-edge fields such as quantum optics\cite{zhang2011optical}, cold molecule research\cite{ni2008high}, atomic clock\cite{boyd2007sr}, and precision spectroscopy\cite{tabosa1991nonlinear}. Its primary advantage lies in its ability to achieve high-density, low-temperature atomic trapping with relatively low technical complexity, providing a stable source of cold atoms for conducting complex experiments.

Rubidium (Rb) and Cesium (Cs), as the most commonly used species in cold atom experiments, are widely employed in studies of cold atomic gases due to their excellent laser cooling properties and rich hyperfine energy level structures \cite{metcalf1999laserca}. Although single-species MOT systems have been extensively optimized, the scope of research has expanded, and this singularity has increasingly become a bottleneck in the study of multicomponent systems. In many emerging fields, such as quantum information processing and cold atom quantum simulation\cite{anand2024dual,zeng2017entangling,cabrera2018quantum}, the need to simultaneously cool and manipulate multiple atomic species has become more prominent. For example, in ultracold heteronuclear molecule synthesis, simultaneous cooling of multiple atomic species is a prerequisite. Since the first synthesis of KRb diatomic molecules through magnetic association and stimulated Raman adiabatic passage in 2008\cite{ni2008high}, heteronuclear molecules such as RbCs, NaK, and NaRb have been successively synthesized\cite{takekoshi2014ultracold,molony2014creation,park2015ultracold,guo2016creation}. Significant progress was made in 2022 with the first quantum-coherent synthesis of triatomic molecules in a laboratory setting\cite{yang2022evidence}. A system capable of cooling multiple atomic species not only enables these complex experiments but also provides new perspectives for studying interatomic interactions, multicomponent effects in mixed cold atomic gases, and dynamical behaviors in diatomic gases\cite{pilch2009observation,burchianti2018dual,catani2008degenerate,delehaye2015critical}.

However, traditional MOT systems are typically designed for specific atomic species from the outset, and the high vacuum environment makes it extremely difficult to add or change atomic species later. To address this issue, we introduced both Rb and Cs atomic sources at the design stage. In principle, the system can simultaneously capture three atomic species: \textsuperscript{85}Rb, \textsuperscript{87}Rb, and \textsuperscript{133}Cs. This design overcomes the limitations of single-species MOT systems, providing a more flexible experimental platform for multicomponent research. Due to the significant frequency difference between the cooling light for Rb and Cs atoms, we designed independent laser systems for each species and achieved precise control to enable efficient cooling and trapping. This technical foundation is a critical guarantee for conducting mixed-species atomic experiments.

\begin{figure*}[htb]
\includegraphics[width=2\columnwidth]{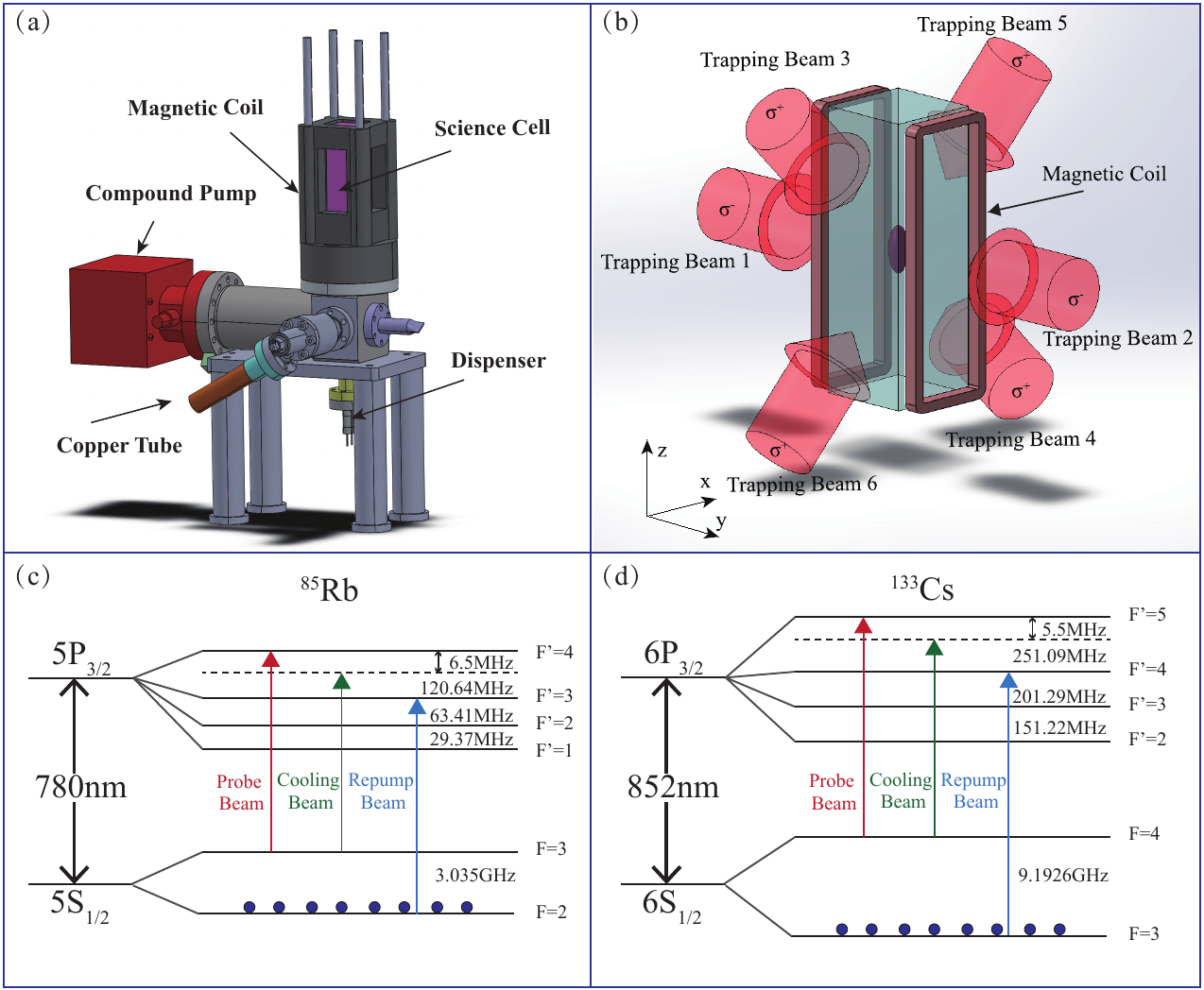}
\caption{ Schematics of the two-species 3D MOT conﬁguration. (a) is a 3D view of the system. (b) is a 3D view of the MOT, only one pair of magnetic field coils is shown. (c) and (d) show the energy level diagrams of the \textsuperscript{85}Rb and \textsuperscript{133}Cs D\textsubscript{2} lines, with the corresponding MOT laser transitions.}
\label{SystemAndEnergyLevel}
\end{figure*}

In this paper, we review and describe a dual-species 3D magneto-optical trap (MOT) apparatus. This system enables simultaneous trapping of \textsuperscript{85}Rb and \textsuperscript{133}Cs atoms at the same position within a vacuum chamber and achieves excellent performance with smaller red detuning of the cooling light. Unlike conventional designs, we arranged the trapping beams at a 45° angle to the vertical axis, rather than perpendicular to the glass cell surface. This configuration ensures complete optical access in the horizontal direction and avoids interference between the cooling and repump beams. Furthermore, with both cooling and repump beams controlled by acousto-optic modulators (AOMs), the MOT operates at a high repetition rate. In our experiments, under the conditions of \textsuperscript{85}Rb trapping light power of 8 mW, repump light power of 4 mW, \textsuperscript{133}Cs trapping light power of 7.5 mW, and repump light power of 1.5 mW, we achieved optical depth (OD) of 3.71 and 3.45 for \textsuperscript{85}Rb and \textsuperscript{133}Cs atoms, respectively. Additionally, we observed suppression of \textsuperscript{133}Cs atom absorption due to \textsuperscript{85}Rb atom clouds, further confirming the high spatial overlap between the two atomic species. In recent years, similar dual-species systems have been applied to many cutting-edge studies, such as the synthesis of heteronuclear Rydberg molecules\cite{peper2021heteronuclear} and the construction of quantum gates\cite{zeng2017entangling,m2023parallel}. Our dual-species MOT apparatus also demonstrates broad potential applications in studies of interspecies interactions and Rydberg molecules. Our system will play a significant role in the intersection of cold atom physics and quantum technology.

\begin{figure*}[htb]
\includegraphics[width=2\columnwidth]{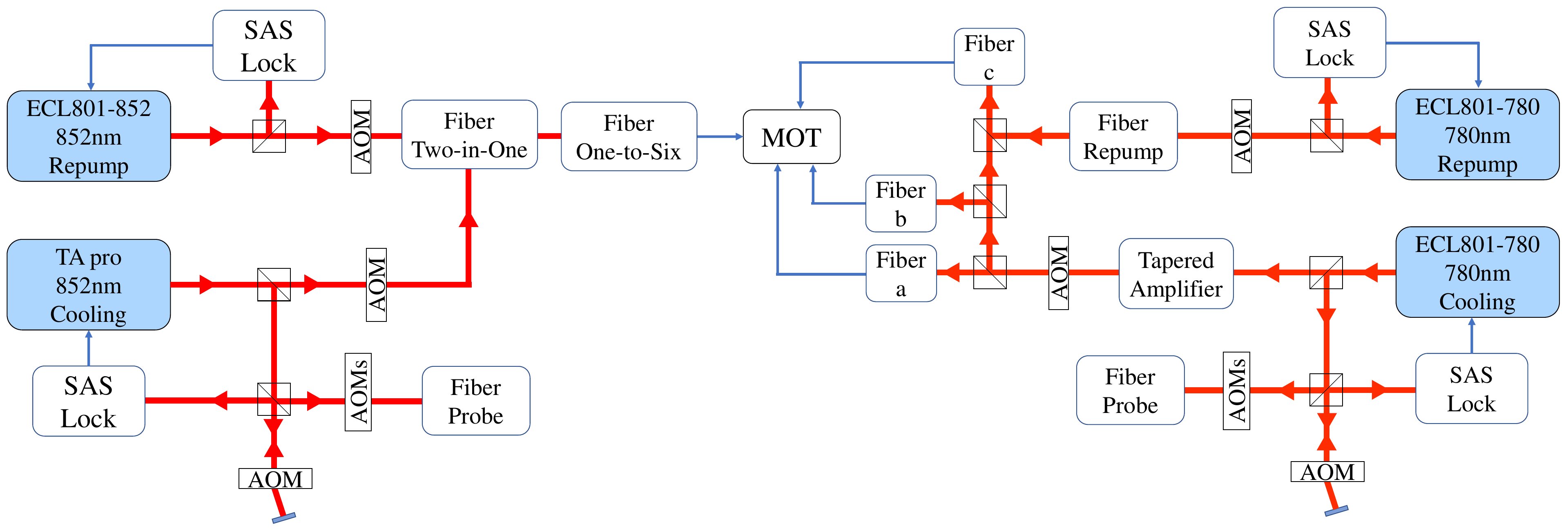}
\caption{A simplified scheme of the laser system for the \textsuperscript{85}Rb-\textsuperscript{133}Cs MOT. All lasers are stabilized to atomic transitions with saturated absorption spectroscopy. The cooling light for Rb is split using Polarizing Beam Splitters (PBS), while the cooling light for Cs is split using a fiber optic beam splitter. Frequency of the beams in different parts of the set-up are controlled by acousto-optic modulators (AOM).}
\label{Net optical}
\end{figure*}

\section{3D MOT setup}

A typical MOT apparatus comprises an ultra-high vacuum (UHV) cell, magnetic fields, and laser beams. The vacuum system we use is shown in the Fig.~\ref{SystemAndEnergyLevel}(a). The rectangular UHV science cell we used is made from 3 mm thick borosilicate glass and has internal dimensions of 20 $\times$ 20 $\times$ 60 mm. A cylindrical graded index section increases the overall length of the cell from the flange to $\sim$ 110 mm. We install totally two Rubidium(Rb) dispensers(SAES,RB/NF/7/25 FT10+10) to provide controllable Rb vapor pressure. During the MOT operation, we only run one dispenser and the other one is reserved for backup. A constant voltage current source is used to power the dispensers, and the Rb atom vapor pressure is controlled by adjusting the current. The naturally occurring Rubidium is composed of \textsuperscript{85}Rb (72.2\%) and \textsuperscript{87}Rb (27.8\%). In this work, we describe the 3D MOT for \textsuperscript{85}Rb atoms, but the system can also be used to cool and trap \textsuperscript{87}Rb atoms by changing the MOT laser frequencies. For the Cs source, we place metallic Cs in a copper tube connected to a valve, which is installed on one side of the multi-port vacuum chamber. A heating tape is wrapped around the copper tube, powered by another constant voltage current source. The Cs vapor pressure is adjusted by changing the valve size and the current supplied to the heating tape. The experiment needs the pressure inside the science cell to be below 10\textsuperscript{-5} Pa. In our system, we first use a mechanical pump to lower the pressure to 10\textsuperscript{-4} Pa. Then, we start the compound pump, which has an ion pump and a getter pump. Once the compound pump is operational, we strategically seal the mechanical pump's interface, allowing its complete removal and optimizing the experimental setup's spatial efficiency. With the help of both pumps, the final pressure in our science cell is below 2.1 $\times$ 10\textsuperscript{-9} Pa.

As illustrated in Fig.~\ref{SystemAndEnergyLevel}(b), the magneto-optical configuration comprises a spatially varying magnetic field produced from a magnetic coil and six trapping beams. To produce the magnetic field, we utilized MAG-3000 magnetic field coils from ColdQuanta\textsuperscript{TM}, which consist of six independently powered coils integrated into a single structure mounted outside the glass vacuum cell. In this system, we only power the two opposing coils in reverse to create the spatial magnetic field. The two powered coils have 120 turns each, measuring 60 mm in length and 35 mm in width, with a spacing of 45 mm between them. In the experiment, the coils are powered at 1.58 A, resulting in an axial magnetic field gradient of approximately 22 Gauss/cm at the center of the trapping area, which meets the experimental requirements. By varying the current, we can adjust the spatial magnetic field gradient to satisfy different experimental conditions. Only the two opposing coils among the six independently powered coils are used to construct the magneto-optical trap for capturing atoms, while the remaining coils can be utilized for compensating the magnetic field and other operations, providing a significant operational redundancy.

The separated laser systems are used for cooling and trapping Rubidium (Rb) and Cesium (Cs), as shown in Fig.~\ref{Net optical}. The relevant energy level diagrams for the cooling processes of the two species are shown in Fig.~\ref{SystemAndEnergyLevel}(c)(d). Both Rubidium and Cesium require separate lasers to provide cooling and repump light due to the relatively large hyperfine splitting in their ground states. All lasers in our system are frequency-locked using saturated absorption spectroscopy. For Rb, the cooling laser is red detuned by 6.5 MHz from the transition $|S_{1/2}, F=3\rangle \to |P_{3/2}, F' = 4\rangle$, and the repump laser is on resonance with the transition $|S_{1/2}, F=2\rangle \to |P_{3/2}, F' = 3\rangle$. For Cs, the cooling laser is red detuned by 5.5 MHz from the transition $|S_{1/2}, F=4\rangle \to |P_{3/2}, F' = 5\rangle$, and the repump laser is on resonance with the transition $|S_{1/2}, F=3\rangle \to |P_{3/2}, F' = 4\rangle$. For the Rb probe light, we choose to split a portion from the cooling light path. The final probe light frequency corresponds to the resonance of the transition $|S_{1/2}, F=3\rangle \to |P_{3/2}, F' = 4\rangle$. The probe light for Cs is also split from the cooling light, following the same procedure as for Rb. The final frequency corresponds to the $|S_{1/2}, F=4\rangle \to |P_{3/2}, F' = 5\rangle$ resonance. All lasers are coupled into optical fibers after passing through the AOM and transmitted to the vicinity of the MOT. The three fibers for Rb and Cs are connected to a beam expander located near the MOT to complete the beam expansion. After passing through a $\lambda$/2 wave plate and a $\lambda$/4 wave plate, the cooling light for both Rb and Cs is combined using a dichroic mirror (DM). The combined beams pass through a broadband $\lambda$/4 wave plate before entering the science cell. One pair of counterpropagating trapping laser beams (1 and 2) are aligned along the y axis. Unlike conventional 3D MOT setups, where the trapping beams are orthogonal to the cell surfaces, in our system, the four trapping beams (labeled 3, 4, 5, and 6) are aligned at 45° angles to both the x and z axes, as shown in Fig.~\ref{SystemAndEnergyLevel}(b). Because this configuration allows full horizontal optical access to the trapped atomic ensembles without spatial interference from the trapping or repump laser beams. During atom trapping, we keep the broadband $\lambda$/4 wave plate stationary and adjust the wave plates after the beam expanders for the Rb and Cs cooling light. This ensures that the cooling light for both Rb and Cs entering the science cell through the broadband $\lambda$/4 wave plate is in the appropriate circular polarization state, which is required for trapping the atoms. The polarizations ($\sigma^{+}$ and $\sigma^{-}$) of the six trapping beams are shown in Fig.~\ref{SystemAndEnergyLevel}(b). In the two-species 3D MOT, we simultaneously trap two types of atoms, using a cooling laser power of only 8 mW for Rb and only 7.5 mW for Cs, along with a repump laser power of 4 mW for Rb and 1.5 mW for Cs.

\section{Optical depth and measurement}

\begin{figure}[t]
    \includegraphics[width=1\linewidth]{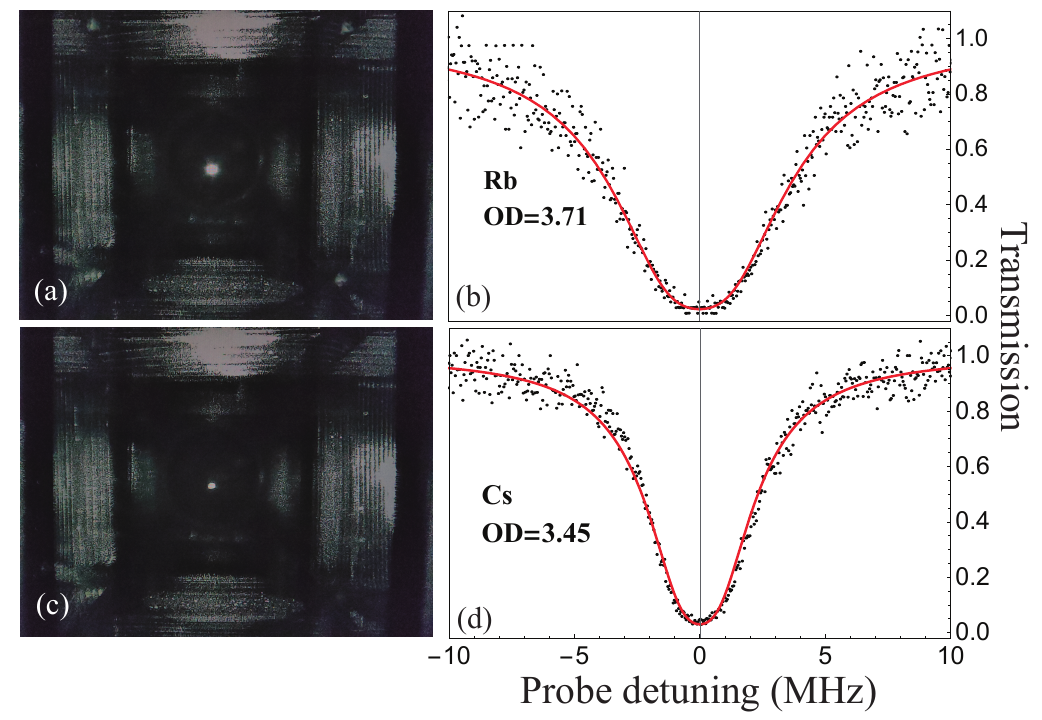}
    \caption{Fluorescence images of trapped atoms in the 3D MOT and typical absorption measurement results. (a) and (b) show the results for Rb atoms. (c) and (d) show the results for Cs atoms. The white point in the center of the image represents the trapped atoms. The circular points are experimental data, and the solid lines are theoretical curves plotted from Eq.~\ref{equation1}. }
    \label{CCD}
\end{figure}

\begin{figure}[htb]
    \includegraphics[width=1\linewidth]{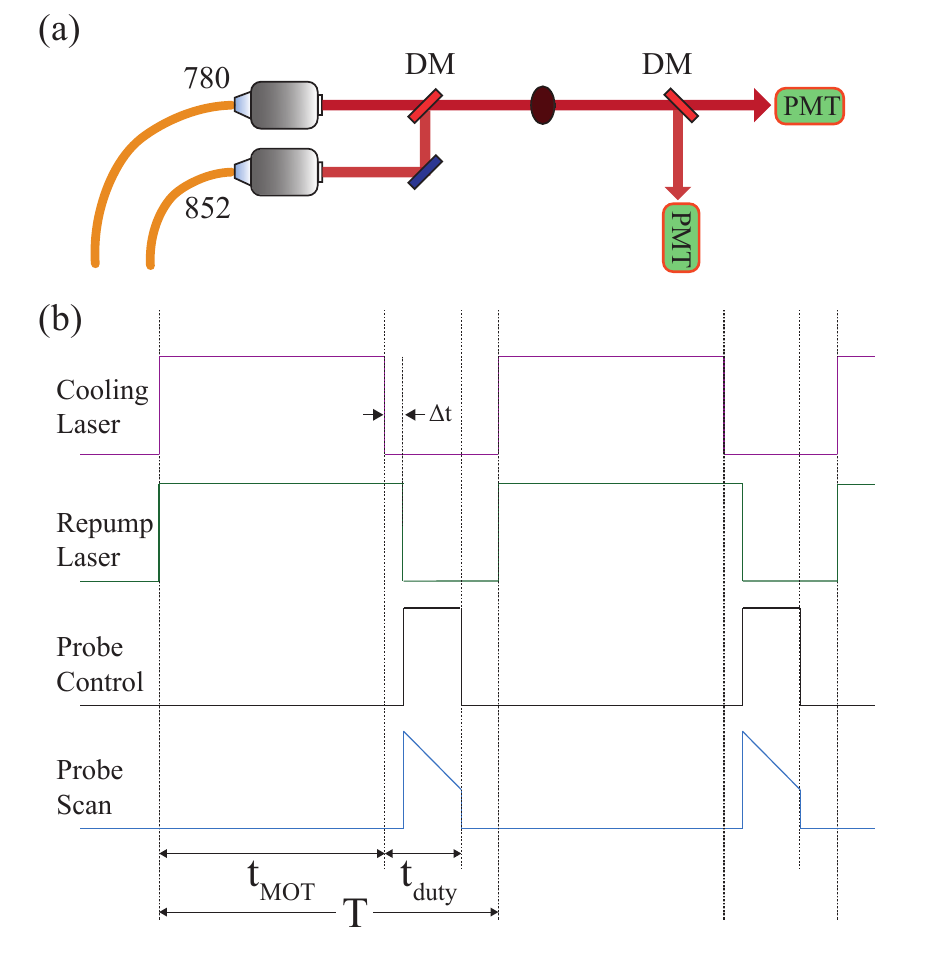}
    \caption{OD measurement scheme. (a) OD measurement optical setup. Two probe light is combined using a dichroic mirror (DM) and pass through the trapped atoms before being detected by the photomultiplier tube (PMT). (b) MOT and OD measurement timing. The total measurement time period T is 10 ms, with the MOT trapping time \(t\textsubscript{MOT}\) = 9 ms and the measure duty time \(t\textsubscript{duty}\) = 0.55 ms.}
    \label{OD}
\end{figure}

The Fig.~\ref{CCD}(a)(c) shows the CCD fluorescence image of the trapped atoms in the 3D MOT. The experimental system employs a design where two atomic capture laser beams are co-injected into the science cell, sharing a common set of magnetic field coils. This configuration allows for the simultaneous capture of \textsuperscript{85}Rb and \textsuperscript{133}Cs atoms at the same spatial location within the cell. Compared to typical systems, the red detuning of the cooling light in this system is significantly lower\cite{harris2008magnetic,witkowski2017dual,isichenko2023photonic,squires2008high}. Generally, the red detuning of the capture laser is approximately 2 $\sim$ 3 $\Gamma$ (where $\Gamma$ is the natural linewidth of the cooling transition). In this system, the red detuning for capturing \textsuperscript{85}Rb atoms is 1.05 $\Gamma$, while for \textsuperscript{133}Cs atoms, it is 1.07 $\Gamma$. We take OD measurement to characterize the 3D MOT properties. Optical depth (OD), as an important physical quantity used to quantify the degree of attenuation of light as it propagates through a medium, helps in understanding and predicting the behavior of light in different media and has wide-ranging applications. The OD measurement scheme is shown in Fig.~\ref{OD}, where (a) is the optical setup, and (b) is the timing. The two probe beams are transmitted via optical fibers to the vicinity of the MOT. After passing through a half-wave plate and a polarizing beam splitter (PBS), the two probe beams are combined using a dichroic mirror (DM). The combined beam is then focused onto the trapped atomic cloud using an achromatic lens with a focal length of 100 mm. After passing through the atomic cloud, the probe light is collimated again using another achromatic lens with the same focal length (100 mm). The beams are then separated by the dichroic mirror and individually coupled into optical fibers. Finally, we measure the probe absorption spectrum with a photomultiplier tube (PMT). The measurement is taken periodically. At each period of T = 10 ms, we set the MOT trapping time \(t\textsubscript{MOT}\) = 9 ms and the measurement duty (including state preparation and OD measurement) time \(t\textsubscript{duty}\) = 0.55 ms. At each cycle, after the cooling laser is switched off, we remain on the repump laser for additional  \(\Delta t\) = 0.5 ms to optically pump all the atoms into the state $|S_{1/2}, F=3\rangle$ for \textsuperscript{85}Rb and $|S_{1/2}, F=4\rangle$ for \textsuperscript{133}Cs , which are preferable for the OD scheme. After the atoms are prepared in the suitable state, we switch on the probe lasers for absorption measurement inside the duty window. All the laser powers are controlled by AOMs.

The transmission of the weak probe laser after passing through the medium is given by \cite{zhang2012dark}

\begin{equation}
\mathrm{Tran}(\omega_{p})=\left| e^{ik(\omega _{p})L} \right|^{2}  =e^{-2\mathrm{Im}[k(\omega _{p})]L},
\label{equation1}
\end{equation}

where \textit{L} is the medium length and $k(\omega_{p})=(\omega_{p}/c)\sqrt{1+\chi}\simeq(\omega_{p}/c)(1+\chi/2)$ is the complex wave number in the medium with the speed of light in vacuum \textit{c}. The linear susceptibility (complex) is given as \cite{wei2009optical}

\begin{equation}
\chi=-\frac{\alpha _{0}}{k_{0}} \frac{\gamma _{13}}{\Delta \omega _{p}+i\gamma_{13}}.
\label{equation2}
\end{equation}

For Rb, $\alpha_{0}$ is the on-resonance absorption coefficient of the transition $|S_{1/2}, F=3\rangle \to |P_{3/2}, F' = 4\rangle$, $k_{0}=\omega_{r}/c$, $\omega_{r}$ is the resonance frequency of the transition $|S_{1/2}, F=3\rangle \to |P_{3/2}, F' = 4\rangle$, $\Delta\omega_{p}=\omega_{p}-\omega_{r}$ is the probe detuning. The atomic-lifetime-determined electric dipole relaxation rate between $|S_{1/2}, F=3\rangle \to |P_{3/2}, F' = 4\rangle$ is $\gamma_{13}$. The atomic optical depth OD = $\alpha_{0}L$ and $\gamma_{13}$ are obtained as the best fitting parameters of Eq.~\ref{equation1} to the measurement data. For Cs, the relevant parameters can be replaced with those specific to Cs atoms during the calculations.

Fig.~\ref{CCD}(b)(d) presents typical absorption measurement results for the two-species 3D magneto-optical trap MOT. The probe lasers are maximally absorbed at resonance, as shown in Fig.~\ref{CCD}(b)(d). The solid curves represent the best fit of Eq.~\ref{equation1} to the experimental data, yielding an optical depth (OD) of 3.71 and $\gamma_{13}$ = 2$\pi$ × 1.8 MHz for \textsuperscript{85}Rb, and an OD of 3.45 and $\gamma_{13}$ = 2$\pi$ × 1.2 MHz for \textsuperscript{133}Cs.

\begin{figure}
    \centering
    \includegraphics[width=0.75\linewidth]{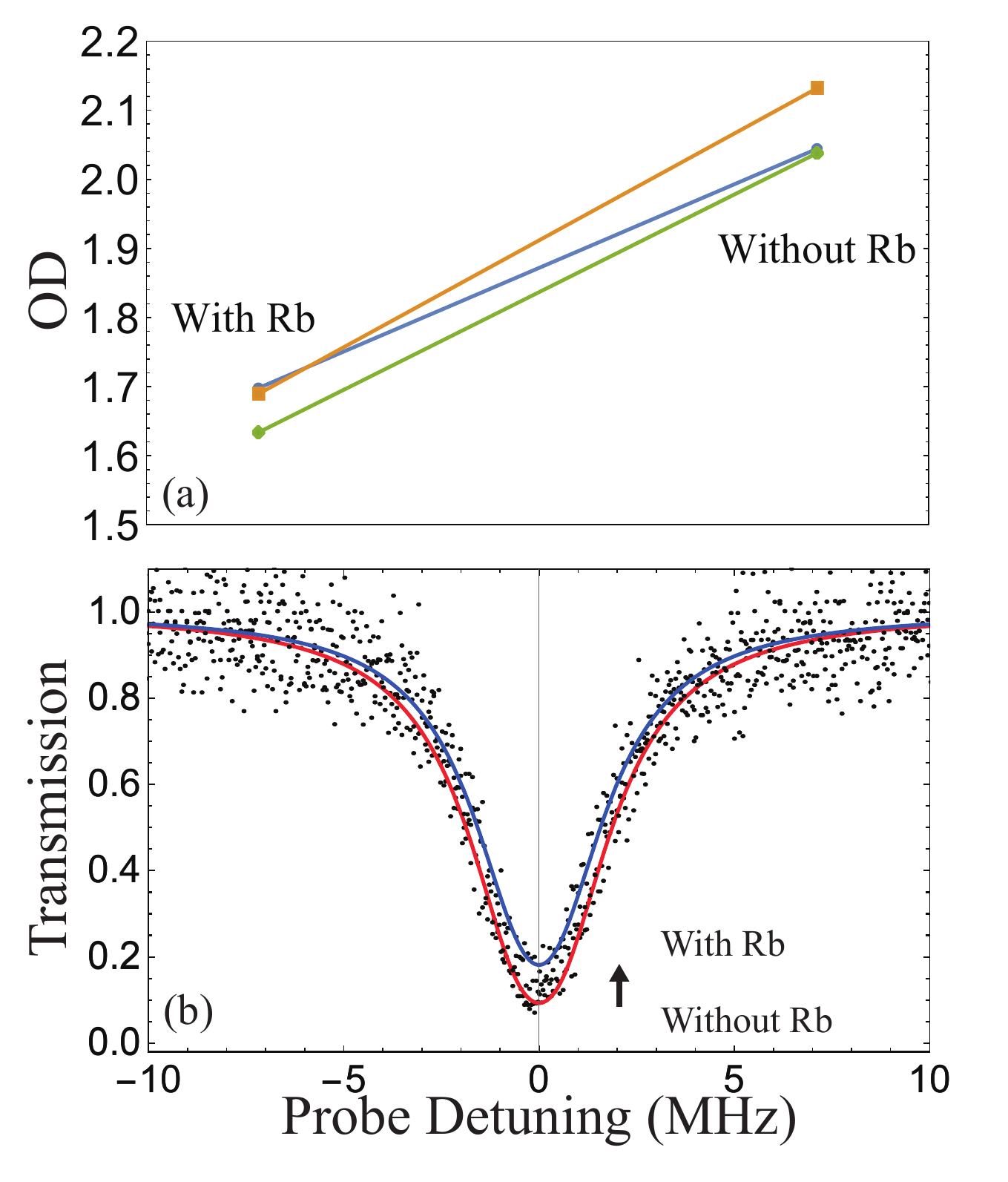}
    \caption{Suppression effect of Rb presence on Cs absorption. (a) Shows repeated measurements, left data points represent Cs OD measurements during co-trapping of Rb and Cs atoms, right data points represent Cs OD measurements in single-species trapping. (b) Comparative OD measurements showing original data points black and fitting curves for Cs atoms with blue and without red Rb atom co-presence.}
    \label{RbDisCs}
\end{figure}

Fig.~\ref{RbDisCs} illustrates the inhibitory effect of trapped Rb atoms on the absorption of trapped Cs atoms when both species are simultaneously captured. Initially, with both Rb and Cs trapping light activated, the OD of the Cs atomic cloud was measured while both species were present in the MOT. Subsequently, the Rb trapping light was turned off, releasing the Rb atomic cloud, and the OD of the Cs atomic cloud was measured again with only Cs atoms trapped. From Fig.~\ref{RbDisCs}, it is evident that the OD of the Cs atomic cloud increased by at least 16\% (from 1.69 to 2.04) in the absence of the Rb atomic cloud, indicating a clear inhibitory effect of the Rb atoms on the absorption of Cs atoms. This effect is attributed to inelastic scattering between the Cs and Rb atoms, which are trapped at the same spatial location, interfering with the loading of Cs atoms and resulting in a reduction of the OD for Cs\cite{harris2008magnetic}. This phenomenon also indicates a significant spatial overlap between the two types of atoms captured in this system.

\section{Conclusion}

In summary, we have described the apparatus of 3D MOT of \textsuperscript{85}Rb and \textsuperscript{133}Cs atoms and its OD measurements. In our six-trapping-beam optical setup, four of the trapping beams are aligned at 45° angles to the vertical axis, unlike the conventional setup where trapping beams are orthogonal to the cell surfaces. This trapping-repump beam configuration opens full optical access along the longitudinal axis and it is important for quantum optics experiments. With a total trapping laser power of 8 mW and repump laser power of 4 mW for \textsuperscript{85}Rb atoms, and a total trapping laser power of 7.5 mW and repump laser power of 1.5 mW for \textsuperscript{133}Cs atoms, we obtain cold atomic ensembles with ODs of 3.71 for \textsuperscript{85}Rb and 3.45 for \textsuperscript{133}Cs. The observed suppression of \textsuperscript{133}Cs atom absorption due to the presence of trapped \textsuperscript{85}Rb atoms indicates a high degree of spatial overlap between the two species in the trapping state. In this work, we describe the 3D MOT for \textsuperscript{85}Rb and \textsuperscript{133}Cs atoms, but the system can also be used to trap \textsuperscript{87}Rb atoms by changing the MOT laser frequencies. 

Future enhancements will focus on refining control schemes, improving trapping efficiency, and incorporating Rydberg state excitation. These upgrades will enable studies of long-range interactions \cite{duspayev2021long,defenu2023long} and heteronuclear Rydberg molecules \cite{bendkowsky2009observation,shaffer2018ultracold}, contributing to the integration of Rydberg physics with ultracold atom research. With its low laser power requirements and capability to simultaneously trap Rb and Cs atoms, our system offers a nearly optimal cold atom source for advanced quantum optics research. Our systems will support research in cutting-edge fields such as quantum computing \cite{saffman2010quantum}, quantum simulation \cite{browaeys2020many,dobrzyniecki2024tunable}, and precision measurement \cite{Ding2022EnhancedMA}.

\bibliography{ref}  % 说明bib文件名称

\begin{thebibliography}{34}%
\makeatletter
\providecommand \@ifxundefined [1]{%
 \@ifx{#1\undefined}
}%
\providecommand \@ifnum [1]{%
 \ifnum #1\expandafter \@firstoftwo
 \else \expandafter \@secondoftwo
 \fi
}%
\providecommand \@ifx [1]{%
 \ifx #1\expandafter \@firstoftwo
 \else \expandafter \@secondoftwo
 \fi
}%
\providecommand \natexlab [1]{#1}%
\providecommand \enquote  [1]{``#1''}%
\providecommand \bibnamefont  [1]{#1}%
\providecommand \bibfnamefont [1]{#1}%
\providecommand \citenamefont [1]{#1}%
\providecommand \href@noop [0]{\@secondoftwo}%
\providecommand \href [0]{\begingroup \@sanitize@url \@href}%
\providecommand \@href[1]{\@@startlink{#1}\@@href}%
\providecommand \@@href[1]{\endgroup#1\@@endlink}%
\providecommand \@sanitize@url [0]{\catcode `\\12\catcode `\$12\catcode `\&12\catcode `\#12\catcode `\^12\catcode `\_12\catcode `\%12\relax}%
\providecommand \@@startlink[1]{}%
\providecommand \@@endlink[0]{}%
\providecommand \url  [0]{\begingroup\@sanitize@url \@url }%
\providecommand \@url [1]{\endgroup\@href {#1}{\urlprefix }}%
\providecommand \urlprefix  [0]{URL }%
\providecommand \Eprint [0]{\href }%
\providecommand \doibase [0]{https://doi.org/}%
\providecommand \selectlanguage [0]{\@gobble}%
\providecommand \bibinfo  [0]{\@secondoftwo}%
\providecommand \bibfield  [0]{\@secondoftwo}%
\providecommand \translation [1]{[#1]}%
\providecommand \BibitemOpen [0]{}%
\providecommand \bibitemStop [0]{}%
\providecommand \bibitemNoStop [0]{.\EOS\space}%
\providecommand \EOS [0]{\spacefactor3000\relax}%
\providecommand \BibitemShut  [1]{\csname bibitem#1\endcsname}%
\let\auto@bib@innerbib\@empty
%</preamble>
\bibitem [{\citenamefont {Raab}\ \emph {et~al.}(1987)\citenamefont {Raab}, \citenamefont {Prentiss}, \citenamefont {Cable}, \citenamefont {Chu},\ and\ \citenamefont {Pritchard}}]{raab1987trapping}%
  \BibitemOpen
  \bibfield  {author} {\bibinfo {author} {\bibfnamefont {E.~L.}\ \bibnamefont {Raab}}, \bibinfo {author} {\bibfnamefont {M.}~\bibnamefont {Prentiss}}, \bibinfo {author} {\bibfnamefont {A.}~\bibnamefont {Cable}}, \bibinfo {author} {\bibfnamefont {S.}~\bibnamefont {Chu}},\ and\ \bibinfo {author} {\bibfnamefont {D.~E.}\ \bibnamefont {Pritchard}},\ }\bibfield  {title} {\bibinfo {title} {Trapping of neutral sodium atoms with radiation pressure},\ }\href {https://journals.aps.org/prl/abstract/10.1103/PhysRevLett.59.2631} {\bibfield  {journal} {\bibinfo  {journal} {Physical review letters}\ }\textbf {\bibinfo {volume} {59}},\ \bibinfo {pages} {2631} (\bibinfo {year} {1987})}\BibitemShut {NoStop}%
\bibitem [{\citenamefont {Zhang}\ \emph {et~al.}(2011)\citenamefont {Zhang}, \citenamefont {Chen}, \citenamefont {Liu}, \citenamefont {Loy}, \citenamefont {Wong},\ and\ \citenamefont {Du}}]{zhang2011optical}%
  \BibitemOpen
  \bibfield  {author} {\bibinfo {author} {\bibfnamefont {S.}~\bibnamefont {Zhang}}, \bibinfo {author} {\bibfnamefont {J.}~\bibnamefont {Chen}}, \bibinfo {author} {\bibfnamefont {C.}~\bibnamefont {Liu}}, \bibinfo {author} {\bibfnamefont {M.}~\bibnamefont {Loy}}, \bibinfo {author} {\bibfnamefont {G.~K.}\ \bibnamefont {Wong}},\ and\ \bibinfo {author} {\bibfnamefont {S.}~\bibnamefont {Du}},\ }\bibfield  {title} {\bibinfo {title} {Optical precursor of a single photon},\ }\href {https://journals.aps.org/prl/abstract/10.1103/PhysRevLett.106.243602} {\bibfield  {journal} {\bibinfo  {journal} {Physical Review Letters}\ }\textbf {\bibinfo {volume} {106}},\ \bibinfo {pages} {243602} (\bibinfo {year} {2011})}\BibitemShut {NoStop}%
\bibitem [{\citenamefont {Ni}\ \emph {et~al.}(2008)\citenamefont {Ni}, \citenamefont {Ospelkaus}, \citenamefont {De~Miranda}, \citenamefont {Pe'Er}, \citenamefont {Neyenhuis}, \citenamefont {Zirbel}, \citenamefont {Kotochigova}, \citenamefont {Julienne}, \citenamefont {Jin},\ and\ \citenamefont {Ye}}]{ni2008high}%
  \BibitemOpen
  \bibfield  {author} {\bibinfo {author} {\bibfnamefont {K.-K.}\ \bibnamefont {Ni}}, \bibinfo {author} {\bibfnamefont {S.}~\bibnamefont {Ospelkaus}}, \bibinfo {author} {\bibfnamefont {M.}~\bibnamefont {De~Miranda}}, \bibinfo {author} {\bibfnamefont {A.}~\bibnamefont {Pe'Er}}, \bibinfo {author} {\bibfnamefont {B.}~\bibnamefont {Neyenhuis}}, \bibinfo {author} {\bibfnamefont {J.}~\bibnamefont {Zirbel}}, \bibinfo {author} {\bibfnamefont {S.}~\bibnamefont {Kotochigova}}, \bibinfo {author} {\bibfnamefont {P.}~\bibnamefont {Julienne}}, \bibinfo {author} {\bibfnamefont {D.}~\bibnamefont {Jin}},\ and\ \bibinfo {author} {\bibfnamefont {J.}~\bibnamefont {Ye}},\ }\bibfield  {title} {\bibinfo {title} {A high phase-space-density gas of polar molecules},\ }\href {https://www.science.org/doi/full/10.1126/science.1163861} {\bibfield  {journal} {\bibinfo  {journal} {science}\ }\textbf {\bibinfo {volume} {322}},\ \bibinfo {pages} {231} (\bibinfo {year} {2008})}\BibitemShut {NoStop}%
\bibitem [{\citenamefont {Boyd}\ \emph {et~al.}(2007)\citenamefont {Boyd}, \citenamefont {Ludlow}, \citenamefont {Blatt}, \citenamefont {Foreman}, \citenamefont {Ido}, \citenamefont {Zelevinsky},\ and\ \citenamefont {Ye}}]{boyd2007sr}%
  \BibitemOpen
  \bibfield  {author} {\bibinfo {author} {\bibfnamefont {M.~M.}\ \bibnamefont {Boyd}}, \bibinfo {author} {\bibfnamefont {A.~D.}\ \bibnamefont {Ludlow}}, \bibinfo {author} {\bibfnamefont {S.}~\bibnamefont {Blatt}}, \bibinfo {author} {\bibfnamefont {S.~M.}\ \bibnamefont {Foreman}}, \bibinfo {author} {\bibfnamefont {T.}~\bibnamefont {Ido}}, \bibinfo {author} {\bibfnamefont {T.}~\bibnamefont {Zelevinsky}},\ and\ \bibinfo {author} {\bibfnamefont {J.}~\bibnamefont {Ye}},\ }\bibfield  {title} {\bibinfo {title} {$^{87}${Sr} lattice clock with inaccuracy below $10^{-15}$},\ }\href {https://journals.aps.org/prl/abstract/10.1103/PhysRevLett.98.083002} {\bibfield  {journal} {\bibinfo  {journal} {Physical Review Letters}\ }\textbf {\bibinfo {volume} {98}},\ \bibinfo {pages} {083002} (\bibinfo {year} {2007})}\BibitemShut {NoStop}%
\bibitem [{\citenamefont {Tabosa}\ \emph {et~al.}(1991)\citenamefont {Tabosa}, \citenamefont {Chen}, \citenamefont {Hu}, \citenamefont {Lee},\ and\ \citenamefont {Kimble}}]{tabosa1991nonlinear}%
  \BibitemOpen
  \bibfield  {author} {\bibinfo {author} {\bibfnamefont {J.}~\bibnamefont {Tabosa}}, \bibinfo {author} {\bibfnamefont {G.}~\bibnamefont {Chen}}, \bibinfo {author} {\bibfnamefont {Z.}~\bibnamefont {Hu}}, \bibinfo {author} {\bibfnamefont {R.}~\bibnamefont {Lee}},\ and\ \bibinfo {author} {\bibfnamefont {H.}~\bibnamefont {Kimble}},\ }\bibfield  {title} {\bibinfo {title} {Nonlinear spectroscopy of cold atoms in a spontaneous-force optical trap},\ }\href {https://journals.aps.org/prl/abstract/10.1103/PhysRevLett.66.3245} {\bibfield  {journal} {\bibinfo  {journal} {Physical review letters}\ }\textbf {\bibinfo {volume} {66}},\ \bibinfo {pages} {3245} (\bibinfo {year} {1991})}\BibitemShut {NoStop}%
\bibitem [{\citenamefont {Metcalf}\ and\ \citenamefont {van~der Straten}(1999)}]{metcalf1999laserca}%
  \BibitemOpen
  \bibfield  {author} {\bibinfo {author} {\bibfnamefont {H.~J.}\ \bibnamefont {Metcalf}}\ and\ \bibinfo {author} {\bibfnamefont {P.}~\bibnamefont {van~der Straten}},\ }\bibfield  {title} {\bibinfo {title} {Laser cooling and trapping},\ }\href {https://api.semanticscholar.org/CorpusID:14942756} {\bibfield  {journal} {\bibinfo  {journal} {Peking University-World Scientific Advanced Physics Series}\ } (\bibinfo {year} {1999})}\BibitemShut {NoStop}%
\bibitem [{\citenamefont {Anand}\ \emph {et~al.}(2024)\citenamefont {Anand}, \citenamefont {Bradley}, \citenamefont {White}, \citenamefont {Ramesh}, \citenamefont {Singh},\ and\ \citenamefont {Bernien}}]{anand2024dual}%
  \BibitemOpen
  \bibfield  {author} {\bibinfo {author} {\bibfnamefont {S.}~\bibnamefont {Anand}}, \bibinfo {author} {\bibfnamefont {C.~E.}\ \bibnamefont {Bradley}}, \bibinfo {author} {\bibfnamefont {R.}~\bibnamefont {White}}, \bibinfo {author} {\bibfnamefont {V.}~\bibnamefont {Ramesh}}, \bibinfo {author} {\bibfnamefont {K.}~\bibnamefont {Singh}},\ and\ \bibinfo {author} {\bibfnamefont {H.}~\bibnamefont {Bernien}},\ }\bibfield  {title} {\bibinfo {title} {A dual-species {R}ydberg array},\ }\href {https://www.nature.com/articles/s41567-024-02638-2} {\bibfield  {journal} {\bibinfo  {journal} {Nature Physics}\ }\textbf {\bibinfo {volume} {20}},\ \bibinfo {pages} {1744} (\bibinfo {year} {2024})}\BibitemShut {NoStop}%
\bibitem [{\citenamefont {Zeng}\ \emph {et~al.}(2017)\citenamefont {Zeng}, \citenamefont {Xu}, \citenamefont {He}, \citenamefont {Liu}, \citenamefont {Liu}, \citenamefont {Wang}, \citenamefont {Papoular}, \citenamefont {Shlyapnikov},\ and\ \citenamefont {Zhan}}]{zeng2017entangling}%
  \BibitemOpen
  \bibfield  {author} {\bibinfo {author} {\bibfnamefont {Y.}~\bibnamefont {Zeng}}, \bibinfo {author} {\bibfnamefont {P.}~\bibnamefont {Xu}}, \bibinfo {author} {\bibfnamefont {X.}~\bibnamefont {He}}, \bibinfo {author} {\bibfnamefont {Y.}~\bibnamefont {Liu}}, \bibinfo {author} {\bibfnamefont {M.}~\bibnamefont {Liu}}, \bibinfo {author} {\bibfnamefont {J.}~\bibnamefont {Wang}}, \bibinfo {author} {\bibfnamefont {D.}~\bibnamefont {Papoular}}, \bibinfo {author} {\bibfnamefont {G.}~\bibnamefont {Shlyapnikov}},\ and\ \bibinfo {author} {\bibfnamefont {M.}~\bibnamefont {Zhan}},\ }\bibfield  {title} {\bibinfo {title} {Entangling two individual atoms of different isotopes via {R}ydberg blockade},\ }\href {https://journals.aps.org/prl/abstract/10.1103/PhysRevLett.119.160502} {\bibfield  {journal} {\bibinfo  {journal} {Physical Review Letters}\ }\textbf {\bibinfo {volume} {119}},\ \bibinfo {pages} {160502} (\bibinfo {year} {2017})}\BibitemShut {NoStop}%
\bibitem [{\citenamefont {Cabrera}\ \emph {et~al.}(2018)\citenamefont {Cabrera}, \citenamefont {Tanzi}, \citenamefont {Sanz}, \citenamefont {Naylor}, \citenamefont {Thomas}, \citenamefont {Cheiney},\ and\ \citenamefont {Tarruell}}]{cabrera2018quantum}%
  \BibitemOpen
  \bibfield  {author} {\bibinfo {author} {\bibfnamefont {C.}~\bibnamefont {Cabrera}}, \bibinfo {author} {\bibfnamefont {L.}~\bibnamefont {Tanzi}}, \bibinfo {author} {\bibfnamefont {J.}~\bibnamefont {Sanz}}, \bibinfo {author} {\bibfnamefont {B.}~\bibnamefont {Naylor}}, \bibinfo {author} {\bibfnamefont {P.}~\bibnamefont {Thomas}}, \bibinfo {author} {\bibfnamefont {P.}~\bibnamefont {Cheiney}},\ and\ \bibinfo {author} {\bibfnamefont {L.}~\bibnamefont {Tarruell}},\ }\bibfield  {title} {\bibinfo {title} {Quantum liquid droplets in a mixture of {B}ose-{E}instein condensates},\ }\href {https://www.science.org/doi/full/10.1126/science.aao5686} {\bibfield  {journal} {\bibinfo  {journal} {Science}\ }\textbf {\bibinfo {volume} {359}},\ \bibinfo {pages} {301} (\bibinfo {year} {2018})}\BibitemShut {NoStop}%
\bibitem [{\citenamefont {Takekoshi}\ \emph {et~al.}(2014)\citenamefont {Takekoshi}, \citenamefont {Reichs{\"o}llner}, \citenamefont {Schindewolf}, \citenamefont {Hutson}, \citenamefont {Le~Sueur}, \citenamefont {Dulieu}, \citenamefont {Ferlaino}, \citenamefont {Grimm},\ and\ \citenamefont {N{\"a}gerl}}]{takekoshi2014ultracold}%
  \BibitemOpen
  \bibfield  {author} {\bibinfo {author} {\bibfnamefont {T.}~\bibnamefont {Takekoshi}}, \bibinfo {author} {\bibfnamefont {L.}~\bibnamefont {Reichs{\"o}llner}}, \bibinfo {author} {\bibfnamefont {A.}~\bibnamefont {Schindewolf}}, \bibinfo {author} {\bibfnamefont {J.~M.}\ \bibnamefont {Hutson}}, \bibinfo {author} {\bibfnamefont {C.~R.}\ \bibnamefont {Le~Sueur}}, \bibinfo {author} {\bibfnamefont {O.}~\bibnamefont {Dulieu}}, \bibinfo {author} {\bibfnamefont {F.}~\bibnamefont {Ferlaino}}, \bibinfo {author} {\bibfnamefont {R.}~\bibnamefont {Grimm}},\ and\ \bibinfo {author} {\bibfnamefont {H.-C.}\ \bibnamefont {N{\"a}gerl}},\ }\bibfield  {title} {\bibinfo {title} {Ultracold dense samples of dipolar {RbCs} molecules in the rovibrational and hyperfine ground state},\ }\href {https://journals.aps.org/prl/abstract/10.1103/PhysRevLett.113.205301} {\bibfield  {journal} {\bibinfo  {journal} {Physical review letters}\ }\textbf {\bibinfo {volume} {113}},\ \bibinfo {pages} {205301} (\bibinfo {year} {2014})}\BibitemShut
  {NoStop}%
\bibitem [{\citenamefont {Molony}\ \emph {et~al.}(2014)\citenamefont {Molony}, \citenamefont {Gregory}, \citenamefont {Ji}, \citenamefont {Lu}, \citenamefont {K{\"o}ppinger}, \citenamefont {Le~Sueur}, \citenamefont {Blackley}, \citenamefont {Hutson},\ and\ \citenamefont {Cornish}}]{molony2014creation}%
  \BibitemOpen
  \bibfield  {author} {\bibinfo {author} {\bibfnamefont {P.~K.}\ \bibnamefont {Molony}}, \bibinfo {author} {\bibfnamefont {P.~D.}\ \bibnamefont {Gregory}}, \bibinfo {author} {\bibfnamefont {Z.}~\bibnamefont {Ji}}, \bibinfo {author} {\bibfnamefont {B.}~\bibnamefont {Lu}}, \bibinfo {author} {\bibfnamefont {M.~P.}\ \bibnamefont {K{\"o}ppinger}}, \bibinfo {author} {\bibfnamefont {C.~R.}\ \bibnamefont {Le~Sueur}}, \bibinfo {author} {\bibfnamefont {C.~L.}\ \bibnamefont {Blackley}}, \bibinfo {author} {\bibfnamefont {J.~M.}\ \bibnamefont {Hutson}},\ and\ \bibinfo {author} {\bibfnamefont {S.~L.}\ \bibnamefont {Cornish}},\ }\bibfield  {title} {\bibinfo {title} {Creation of ultracold $^{87}\mathrm{Rb}^{133}\mathrm{Cs}$ molecules in the rovibrational ground state},\ }\href {https://journals.aps.org/prl/abstract/10.1103/PhysRevLett.113.255301} {\bibfield  {journal} {\bibinfo  {journal} {Physical review letters}\ }\textbf {\bibinfo {volume} {113}},\ \bibinfo {pages} {255301} (\bibinfo {year} {2014})}\BibitemShut {NoStop}%
\bibitem [{\citenamefont {Park}\ \emph {et~al.}(2015)\citenamefont {Park}, \citenamefont {Will},\ and\ \citenamefont {Zwierlein}}]{park2015ultracold}%
  \BibitemOpen
  \bibfield  {author} {\bibinfo {author} {\bibfnamefont {J.~W.}\ \bibnamefont {Park}}, \bibinfo {author} {\bibfnamefont {S.~A.}\ \bibnamefont {Will}},\ and\ \bibinfo {author} {\bibfnamefont {M.~W.}\ \bibnamefont {Zwierlein}},\ }\bibfield  {title} {\bibinfo {title} {Ultracold dipolar gas of fermionic $^{23}\mathrm{Na}^{40}\mathrm{K}$ molecules in their absolute ground state},\ }\href {https://journals.aps.org/prl/abstract/10.1103/PhysRevLett.114.205302} {\bibfield  {journal} {\bibinfo  {journal} {Physical review letters}\ }\textbf {\bibinfo {volume} {114}},\ \bibinfo {pages} {205302} (\bibinfo {year} {2015})}\BibitemShut {NoStop}%
\bibitem [{\citenamefont {Guo}\ \emph {et~al.}(2016)\citenamefont {Guo}, \citenamefont {Zhu}, \citenamefont {Lu}, \citenamefont {Ye}, \citenamefont {Wang}, \citenamefont {Vexiau}, \citenamefont {Bouloufa-Maafa}, \citenamefont {Qu{\'e}m{\'e}ner}, \citenamefont {Dulieu},\ and\ \citenamefont {Wang}}]{guo2016creation}%
  \BibitemOpen
  \bibfield  {author} {\bibinfo {author} {\bibfnamefont {M.}~\bibnamefont {Guo}}, \bibinfo {author} {\bibfnamefont {B.}~\bibnamefont {Zhu}}, \bibinfo {author} {\bibfnamefont {B.}~\bibnamefont {Lu}}, \bibinfo {author} {\bibfnamefont {X.}~\bibnamefont {Ye}}, \bibinfo {author} {\bibfnamefont {F.}~\bibnamefont {Wang}}, \bibinfo {author} {\bibfnamefont {R.}~\bibnamefont {Vexiau}}, \bibinfo {author} {\bibfnamefont {N.}~\bibnamefont {Bouloufa-Maafa}}, \bibinfo {author} {\bibfnamefont {G.}~\bibnamefont {Qu{\'e}m{\'e}ner}}, \bibinfo {author} {\bibfnamefont {O.}~\bibnamefont {Dulieu}},\ and\ \bibinfo {author} {\bibfnamefont {D.}~\bibnamefont {Wang}},\ }\bibfield  {title} {\bibinfo {title} {Creation of an ultracold gas of ground-state dipolar $^{23}\mathrm{Na}^{87}\mathrm{Rb}$ molecules},\ }\href {https://journals.aps.org/prl/abstract/10.1103/PhysRevLett.116.205303} {\bibfield  {journal} {\bibinfo  {journal} {Physical review letters}\ }\textbf {\bibinfo {volume} {116}},\ \bibinfo {pages} {205303} (\bibinfo {year}
  {2016})}\BibitemShut {NoStop}%
\bibitem [{\citenamefont {Yang}\ \emph {et~al.}(2022)\citenamefont {Yang}, \citenamefont {Wang}, \citenamefont {Su}, \citenamefont {Cao}, \citenamefont {Zhang}, \citenamefont {Rui}, \citenamefont {Zhao}, \citenamefont {Bai},\ and\ \citenamefont {Pan}}]{yang2022evidence}%
  \BibitemOpen
  \bibfield  {author} {\bibinfo {author} {\bibfnamefont {H.}~\bibnamefont {Yang}}, \bibinfo {author} {\bibfnamefont {X.-Y.}\ \bibnamefont {Wang}}, \bibinfo {author} {\bibfnamefont {Z.}~\bibnamefont {Su}}, \bibinfo {author} {\bibfnamefont {J.}~\bibnamefont {Cao}}, \bibinfo {author} {\bibfnamefont {D.-C.}\ \bibnamefont {Zhang}}, \bibinfo {author} {\bibfnamefont {J.}~\bibnamefont {Rui}}, \bibinfo {author} {\bibfnamefont {B.}~\bibnamefont {Zhao}}, \bibinfo {author} {\bibfnamefont {C.-L.}\ \bibnamefont {Bai}},\ and\ \bibinfo {author} {\bibfnamefont {J.-W.}\ \bibnamefont {Pan}},\ }\bibfield  {title} {\bibinfo {title} {Evidence for the association of triatomic molecules in ultracold $^{23}\mathrm{Na}^{40}\mathrm{K}+^{40}\mathrm{K}$ mixtures},\ }\href {https://www.nature.com/articles/s41586-021-04297-2} {\bibfield  {journal} {\bibinfo  {journal} {Nature}\ }\textbf {\bibinfo {volume} {602}},\ \bibinfo {pages} {229} (\bibinfo {year} {2022})}\BibitemShut {NoStop}%
\bibitem [{\citenamefont {Pilch}\ \emph {et~al.}(2009)\citenamefont {Pilch}, \citenamefont {Lange}, \citenamefont {Prantner}, \citenamefont {Kerner}, \citenamefont {Ferlaino}, \citenamefont {N{\"a}gerl},\ and\ \citenamefont {Grimm}}]{pilch2009observation}%
  \BibitemOpen
  \bibfield  {author} {\bibinfo {author} {\bibfnamefont {K.}~\bibnamefont {Pilch}}, \bibinfo {author} {\bibfnamefont {A.}~\bibnamefont {Lange}}, \bibinfo {author} {\bibfnamefont {A.}~\bibnamefont {Prantner}}, \bibinfo {author} {\bibfnamefont {G.}~\bibnamefont {Kerner}}, \bibinfo {author} {\bibfnamefont {F.}~\bibnamefont {Ferlaino}}, \bibinfo {author} {\bibfnamefont {H.-C.}\ \bibnamefont {N{\"a}gerl}},\ and\ \bibinfo {author} {\bibfnamefont {R.}~\bibnamefont {Grimm}},\ }\bibfield  {title} {\bibinfo {title} {Observation of interspecies feshbach resonances in an ultracold {R}b-{C}s mixture},\ }\href {https://journals.aps.org/pra/abstract/10.1103/PhysRevA.79.042718} {\bibfield  {journal} {\bibinfo  {journal} {Physical Review A—Atomic, Molecular, and Optical Physics}\ }\textbf {\bibinfo {volume} {79}},\ \bibinfo {pages} {042718} (\bibinfo {year} {2009})}\BibitemShut {NoStop}%
\bibitem [{\citenamefont {Burchianti}\ \emph {et~al.}(2018)\citenamefont {Burchianti}, \citenamefont {D'Errico}, \citenamefont {Rosi}, \citenamefont {Simoni}, \citenamefont {Modugno}, \citenamefont {Fort},\ and\ \citenamefont {Minardi}}]{burchianti2018dual}%
  \BibitemOpen
  \bibfield  {author} {\bibinfo {author} {\bibfnamefont {A.}~\bibnamefont {Burchianti}}, \bibinfo {author} {\bibfnamefont {C.}~\bibnamefont {D'Errico}}, \bibinfo {author} {\bibfnamefont {S.}~\bibnamefont {Rosi}}, \bibinfo {author} {\bibfnamefont {A.}~\bibnamefont {Simoni}}, \bibinfo {author} {\bibfnamefont {M.}~\bibnamefont {Modugno}}, \bibinfo {author} {\bibfnamefont {C.}~\bibnamefont {Fort}},\ and\ \bibinfo {author} {\bibfnamefont {F.}~\bibnamefont {Minardi}},\ }\bibfield  {title} {\bibinfo {title} {Dual-species {B}ose-{E}instein condensate of $^{41}${K} and $^{87}${R}b in a hybrid trap},\ }\href {https://journals.aps.org/pra/abstract/10.1103/PhysRevA.98.063616} {\bibfield  {journal} {\bibinfo  {journal} {Physical Review A}\ }\textbf {\bibinfo {volume} {98}},\ \bibinfo {pages} {063616} (\bibinfo {year} {2018})}\BibitemShut {NoStop}%
\bibitem [{\citenamefont {Catani}\ \emph {et~al.}(2008)\citenamefont {Catani}, \citenamefont {De~Sarlo}, \citenamefont {Barontini}, \citenamefont {Minardi},\ and\ \citenamefont {Inguscio}}]{catani2008degenerate}%
  \BibitemOpen
  \bibfield  {author} {\bibinfo {author} {\bibfnamefont {J.}~\bibnamefont {Catani}}, \bibinfo {author} {\bibfnamefont {L.}~\bibnamefont {De~Sarlo}}, \bibinfo {author} {\bibfnamefont {G.}~\bibnamefont {Barontini}}, \bibinfo {author} {\bibfnamefont {F.}~\bibnamefont {Minardi}},\ and\ \bibinfo {author} {\bibfnamefont {M.}~\bibnamefont {Inguscio}},\ }\bibfield  {title} {\bibinfo {title} {Degenerate {B}ose-{B}ose mixture in a three-dimensional optical lattice},\ }\href {https://journals.aps.org/pra/abstract/10.1103/PhysRevA.77.011603} {\bibfield  {journal} {\bibinfo  {journal} {Physical Review A—Atomic, Molecular, and Optical Physics}\ }\textbf {\bibinfo {volume} {77}},\ \bibinfo {pages} {011603} (\bibinfo {year} {2008})}\BibitemShut {NoStop}%
\bibitem [{\citenamefont {Delehaye}\ \emph {et~al.}(2015)\citenamefont {Delehaye}, \citenamefont {Laurent}, \citenamefont {Ferrier-Barbut}, \citenamefont {Jin}, \citenamefont {Chevy},\ and\ \citenamefont {Salomon}}]{delehaye2015critical}%
  \BibitemOpen
  \bibfield  {author} {\bibinfo {author} {\bibfnamefont {M.}~\bibnamefont {Delehaye}}, \bibinfo {author} {\bibfnamefont {S.}~\bibnamefont {Laurent}}, \bibinfo {author} {\bibfnamefont {I.}~\bibnamefont {Ferrier-Barbut}}, \bibinfo {author} {\bibfnamefont {S.}~\bibnamefont {Jin}}, \bibinfo {author} {\bibfnamefont {F.}~\bibnamefont {Chevy}},\ and\ \bibinfo {author} {\bibfnamefont {C.}~\bibnamefont {Salomon}},\ }\bibfield  {title} {\bibinfo {title} {Critical velocity and dissipation of an ultracold {B}ose-{F}ermi counterflow},\ }\href {https://journals.aps.org/prl/abstract/10.1103/PhysRevLett.115.265303} {\bibfield  {journal} {\bibinfo  {journal} {Physical review letters}\ }\textbf {\bibinfo {volume} {115}},\ \bibinfo {pages} {265303} (\bibinfo {year} {2015})}\BibitemShut {NoStop}%
\bibitem [{\citenamefont {Peper}\ and\ \citenamefont {Deiglmayr}(2021)}]{peper2021heteronuclear}%
  \BibitemOpen
  \bibfield  {author} {\bibinfo {author} {\bibfnamefont {M.}~\bibnamefont {Peper}}\ and\ \bibinfo {author} {\bibfnamefont {J.}~\bibnamefont {Deiglmayr}},\ }\bibfield  {title} {\bibinfo {title} {Heteronuclear long-range {R}ydberg molecules},\ }\href {https://journals.aps.org/prl/abstract/10.1103/PhysRevLett.126.013001} {\bibfield  {journal} {\bibinfo  {journal} {Physical Review Letters}\ }\textbf {\bibinfo {volume} {126}},\ \bibinfo {pages} {013001} (\bibinfo {year} {2021})}\BibitemShut {NoStop}%
\bibitem [{\citenamefont {M.~Farouk}\ \emph {et~al.}(2023)\citenamefont {M.~Farouk}, \citenamefont {Beterov}, \citenamefont {Xu}, \citenamefont {Bergamini},\ and\ \citenamefont {Ryabtsev}}]{m2023parallel}%
  \BibitemOpen
  \bibfield  {author} {\bibinfo {author} {\bibfnamefont {A.}~\bibnamefont {M.~Farouk}}, \bibinfo {author} {\bibfnamefont {I.~I.}\ \bibnamefont {Beterov}}, \bibinfo {author} {\bibfnamefont {P.}~\bibnamefont {Xu}}, \bibinfo {author} {\bibfnamefont {S.}~\bibnamefont {Bergamini}},\ and\ \bibinfo {author} {\bibfnamefont {I.~I.}\ \bibnamefont {Ryabtsev}},\ }\bibfield  {title} {\bibinfo {title} {Parallel implementation of {CNOT}$^{N}$ and {C}$_{2}${NOT}$^{2}$ gates via homonuclear and heteronuclear f{\"o}rster interactions of {R}ydberg atoms},\ }\href {https://www.mdpi.com/2304-6732/10/11/1280} {\bibfield  {journal} {\bibinfo  {journal} {Photonics}\ }\textbf {\bibinfo {volume} {10}},\ \bibinfo {pages} {1280} (\bibinfo {year} {2023})}\BibitemShut {NoStop}%
\bibitem [{\citenamefont {Harris}\ \emph {et~al.}(2008)\citenamefont {Harris}, \citenamefont {Tierney},\ and\ \citenamefont {Cornish}}]{harris2008magnetic}%
  \BibitemOpen
  \bibfield  {author} {\bibinfo {author} {\bibfnamefont {M.}~\bibnamefont {Harris}}, \bibinfo {author} {\bibfnamefont {P.}~\bibnamefont {Tierney}},\ and\ \bibinfo {author} {\bibfnamefont {S.}~\bibnamefont {Cornish}},\ }\bibfield  {title} {\bibinfo {title} {Magnetic trapping of a cold {R}b--{C}s atomic mixture},\ }\href {https://iopscience.iop.org/article/10.1088/0953-4075/41/3/035303/meta} {\bibfield  {journal} {\bibinfo  {journal} {Journal of physics B: atomic, molecular and optical physics}\ }\textbf {\bibinfo {volume} {41}},\ \bibinfo {pages} {035303} (\bibinfo {year} {2008})}\BibitemShut {NoStop}%
\bibitem [{\citenamefont {Witkowski}\ \emph {et~al.}(2017)\citenamefont {Witkowski}, \citenamefont {Nag{\'o}rny}, \citenamefont {Munoz-Rodriguez}, \citenamefont {Ciury{\l}o}, \citenamefont {{\.Z}uchowski}, \citenamefont {Bilicki}, \citenamefont {Piotrowski}, \citenamefont {Morzy{\'n}ski},\ and\ \citenamefont {Zawada}}]{witkowski2017dual}%
  \BibitemOpen
  \bibfield  {author} {\bibinfo {author} {\bibfnamefont {M.}~\bibnamefont {Witkowski}}, \bibinfo {author} {\bibfnamefont {B.}~\bibnamefont {Nag{\'o}rny}}, \bibinfo {author} {\bibfnamefont {R.}~\bibnamefont {Munoz-Rodriguez}}, \bibinfo {author} {\bibfnamefont {R.}~\bibnamefont {Ciury{\l}o}}, \bibinfo {author} {\bibfnamefont {P.~S.}\ \bibnamefont {{\.Z}uchowski}}, \bibinfo {author} {\bibfnamefont {S.}~\bibnamefont {Bilicki}}, \bibinfo {author} {\bibfnamefont {M.}~\bibnamefont {Piotrowski}}, \bibinfo {author} {\bibfnamefont {P.}~\bibnamefont {Morzy{\'n}ski}},\ and\ \bibinfo {author} {\bibfnamefont {M.}~\bibnamefont {Zawada}},\ }\bibfield  {title} {\bibinfo {title} {Dual {H}g-{R}b magneto-optical trap},\ }\href {https://opg.optica.org/oe/fulltext.cfm?uri=oe-25-4-3165&id=359856} {\bibfield  {journal} {\bibinfo  {journal} {Optics Express}\ }\textbf {\bibinfo {volume} {25}},\ \bibinfo {pages} {3165} (\bibinfo {year} {2017})}\BibitemShut {NoStop}%
\bibitem [{\citenamefont {Isichenko}\ \emph {et~al.}(2023)\citenamefont {Isichenko}, \citenamefont {Chauhan}, \citenamefont {Bose}, \citenamefont {Wang}, \citenamefont {Kunz},\ and\ \citenamefont {Blumenthal}}]{isichenko2023photonic}%
  \BibitemOpen
  \bibfield  {author} {\bibinfo {author} {\bibfnamefont {A.}~\bibnamefont {Isichenko}}, \bibinfo {author} {\bibfnamefont {N.}~\bibnamefont {Chauhan}}, \bibinfo {author} {\bibfnamefont {D.}~\bibnamefont {Bose}}, \bibinfo {author} {\bibfnamefont {J.}~\bibnamefont {Wang}}, \bibinfo {author} {\bibfnamefont {P.~D.}\ \bibnamefont {Kunz}},\ and\ \bibinfo {author} {\bibfnamefont {D.~J.}\ \bibnamefont {Blumenthal}},\ }\bibfield  {title} {\bibinfo {title} {Photonic integrated beam delivery for a rubidium 3{D} magneto-optical trap},\ }\href {https://www.nature.com/articles/s41467-023-38818-6} {\bibfield  {journal} {\bibinfo  {journal} {Nature communications}\ }\textbf {\bibinfo {volume} {14}},\ \bibinfo {pages} {3080} (\bibinfo {year} {2023})}\BibitemShut {NoStop}%
\bibitem [{\citenamefont {Squires}(2008)}]{squires2008high}%
  \BibitemOpen
  \bibfield  {author} {\bibinfo {author} {\bibfnamefont {M.~B.}\ \bibnamefont {Squires}},\ }\emph {\bibinfo {title} {High repetition rate {B}ose-{E}instein condensate production in a compact, transportable vacuum system}},\ \href {https://www.proquest.com/docview/304621157?pq-origsite=gscholar&fromopenview=true&sourcetype=Dissertations%20&%20Theses} {Ph.D. thesis},\ \bibinfo  {school} {University of Colorado at Boulder} (\bibinfo {year} {2008})\BibitemShut {NoStop}%
\bibitem [{\citenamefont {Zhang}\ \emph {et~al.}(2012)\citenamefont {Zhang}, \citenamefont {Chen}, \citenamefont {Liu}, \citenamefont {Zhou}, \citenamefont {Loy}, \citenamefont {Wong},\ and\ \citenamefont {Du}}]{zhang2012dark}%
  \BibitemOpen
  \bibfield  {author} {\bibinfo {author} {\bibfnamefont {S.}~\bibnamefont {Zhang}}, \bibinfo {author} {\bibfnamefont {J.}~\bibnamefont {Chen}}, \bibinfo {author} {\bibfnamefont {C.}~\bibnamefont {Liu}}, \bibinfo {author} {\bibfnamefont {S.}~\bibnamefont {Zhou}}, \bibinfo {author} {\bibfnamefont {M.}~\bibnamefont {Loy}}, \bibinfo {author} {\bibfnamefont {G.~K.~L.}\ \bibnamefont {Wong}},\ and\ \bibinfo {author} {\bibfnamefont {S.}~\bibnamefont {Du}},\ }\bibfield  {title} {\bibinfo {title} {A dark-line two-dimensional magneto-optical trap of $^{85}\mathrm{Rb}$ atoms with high optical depth},\ }\href {https://pubs.aip.org/aip/rsi/article/83/7/073102/354527} {\bibfield  {journal} {\bibinfo  {journal} {Review of Scientific Instruments}\ }\textbf {\bibinfo {volume} {83}},\ \bibinfo {pages} {073102} (\bibinfo {year} {2012})}\BibitemShut {NoStop}%
\bibitem [{\citenamefont {Wei}\ \emph {et~al.}(2009)\citenamefont {Wei}, \citenamefont {Chen}, \citenamefont {Loy}, \citenamefont {Wong},\ and\ \citenamefont {Du}}]{wei2009optical}%
  \BibitemOpen
  \bibfield  {author} {\bibinfo {author} {\bibfnamefont {D.}~\bibnamefont {Wei}}, \bibinfo {author} {\bibfnamefont {J.}~\bibnamefont {Chen}}, \bibinfo {author} {\bibfnamefont {M.}~\bibnamefont {Loy}}, \bibinfo {author} {\bibfnamefont {G.~K.}\ \bibnamefont {Wong}},\ and\ \bibinfo {author} {\bibfnamefont {S.}~\bibnamefont {Du}},\ }\bibfield  {title} {\bibinfo {title} {Optical precursors with electromagnetically induced transparency in cold atoms},\ }\href {https://journals.aps.org/prl/abstract/10.1103/PhysRevLett.103.093602} {\bibfield  {journal} {\bibinfo  {journal} {Physical review letters}\ }\textbf {\bibinfo {volume} {103}},\ \bibinfo {pages} {093602} (\bibinfo {year} {2009})}\BibitemShut {NoStop}%
\bibitem [{\citenamefont {Duspayev}\ \emph {et~al.}(2021)\citenamefont {Duspayev}, \citenamefont {Han}, \citenamefont {Viray}, \citenamefont {Ma}, \citenamefont {Zhao},\ and\ \citenamefont {Raithel}}]{duspayev2021long}%
  \BibitemOpen
  \bibfield  {author} {\bibinfo {author} {\bibfnamefont {A.}~\bibnamefont {Duspayev}}, \bibinfo {author} {\bibfnamefont {X.}~\bibnamefont {Han}}, \bibinfo {author} {\bibfnamefont {M.}~\bibnamefont {Viray}}, \bibinfo {author} {\bibfnamefont {L.}~\bibnamefont {Ma}}, \bibinfo {author} {\bibfnamefont {J.}~\bibnamefont {Zhao}},\ and\ \bibinfo {author} {\bibfnamefont {G.}~\bibnamefont {Raithel}},\ }\bibfield  {title} {\bibinfo {title} {Long-range {R}ydberg-atom--ion molecules of {R}b and {C}s},\ }\href {https://journals.aps.org/prresearch/abstract/10.1103/PhysRevResearch.3.023114} {\bibfield  {journal} {\bibinfo  {journal} {Physical Review Research}\ }\textbf {\bibinfo {volume} {3}},\ \bibinfo {pages} {023114} (\bibinfo {year} {2021})}\BibitemShut {NoStop}%
\bibitem [{\citenamefont {Defenu}\ \emph {et~al.}(2023)\citenamefont {Defenu}, \citenamefont {Donner}, \citenamefont {Macr{\`\i}}, \citenamefont {Pagano}, \citenamefont {Ruffo},\ and\ \citenamefont {Trombettoni}}]{defenu2023long}%
  \BibitemOpen
  \bibfield  {author} {\bibinfo {author} {\bibfnamefont {N.}~\bibnamefont {Defenu}}, \bibinfo {author} {\bibfnamefont {T.}~\bibnamefont {Donner}}, \bibinfo {author} {\bibfnamefont {T.}~\bibnamefont {Macr{\`\i}}}, \bibinfo {author} {\bibfnamefont {G.}~\bibnamefont {Pagano}}, \bibinfo {author} {\bibfnamefont {S.}~\bibnamefont {Ruffo}},\ and\ \bibinfo {author} {\bibfnamefont {A.}~\bibnamefont {Trombettoni}},\ }\bibfield  {title} {\bibinfo {title} {Long-range interacting quantum systems},\ }\href {https://journals.aps.org/rmp/abstract/10.1103/RevModPhys.95.035002} {\bibfield  {journal} {\bibinfo  {journal} {Reviews of Modern Physics}\ }\textbf {\bibinfo {volume} {95}},\ \bibinfo {pages} {035002} (\bibinfo {year} {2023})}\BibitemShut {NoStop}%
\bibitem [{\citenamefont {Bendkowsky}\ \emph {et~al.}(2009)\citenamefont {Bendkowsky}, \citenamefont {Butscher}, \citenamefont {Nipper}, \citenamefont {Shaffer}, \citenamefont {L{\"o}w},\ and\ \citenamefont {Pfau}}]{bendkowsky2009observation}%
  \BibitemOpen
  \bibfield  {author} {\bibinfo {author} {\bibfnamefont {V.}~\bibnamefont {Bendkowsky}}, \bibinfo {author} {\bibfnamefont {B.}~\bibnamefont {Butscher}}, \bibinfo {author} {\bibfnamefont {J.}~\bibnamefont {Nipper}}, \bibinfo {author} {\bibfnamefont {J.~P.}\ \bibnamefont {Shaffer}}, \bibinfo {author} {\bibfnamefont {R.}~\bibnamefont {L{\"o}w}},\ and\ \bibinfo {author} {\bibfnamefont {T.}~\bibnamefont {Pfau}},\ }\bibfield  {title} {\bibinfo {title} {Observation of ultralong-range {R}ydberg molecules},\ }\href {https://www.nature.com/articles/nature07945} {\bibfield  {journal} {\bibinfo  {journal} {Nature}\ }\textbf {\bibinfo {volume} {458}},\ \bibinfo {pages} {1005} (\bibinfo {year} {2009})}\BibitemShut {NoStop}%
\bibitem [{\citenamefont {Shaffer}\ \emph {et~al.}(2018)\citenamefont {Shaffer}, \citenamefont {Rittenhouse},\ and\ \citenamefont {Sadeghpour}}]{shaffer2018ultracold}%
  \BibitemOpen
  \bibfield  {author} {\bibinfo {author} {\bibfnamefont {J.}~\bibnamefont {Shaffer}}, \bibinfo {author} {\bibfnamefont {S.}~\bibnamefont {Rittenhouse}},\ and\ \bibinfo {author} {\bibfnamefont {H.}~\bibnamefont {Sadeghpour}},\ }\bibfield  {title} {\bibinfo {title} {Ultracold {R}ydberg molecules},\ }\href {https://www.nature.com/articles/s41467-018-04135-6} {\bibfield  {journal} {\bibinfo  {journal} {Nature communications}\ }\textbf {\bibinfo {volume} {9}},\ \bibinfo {pages} {1965} (\bibinfo {year} {2018})}\BibitemShut {NoStop}%
\bibitem [{\citenamefont {Saffman}\ \emph {et~al.}(2010)\citenamefont {Saffman}, \citenamefont {Walker},\ and\ \citenamefont {M{\o}lmer}}]{saffman2010quantum}%
  \BibitemOpen
  \bibfield  {author} {\bibinfo {author} {\bibfnamefont {M.}~\bibnamefont {Saffman}}, \bibinfo {author} {\bibfnamefont {T.~G.}\ \bibnamefont {Walker}},\ and\ \bibinfo {author} {\bibfnamefont {K.}~\bibnamefont {M{\o}lmer}},\ }\bibfield  {title} {\bibinfo {title} {Quantum information with {R}ydberg atoms},\ }\href {https://journals.aps.org/rmp/abstract/10.1103/RevModPhys.82.2313} {\bibfield  {journal} {\bibinfo  {journal} {Reviews of modern physics}\ }\textbf {\bibinfo {volume} {82}},\ \bibinfo {pages} {2313} (\bibinfo {year} {2010})}\BibitemShut {NoStop}%
\bibitem [{\citenamefont {Browaeys}\ and\ \citenamefont {Lahaye}(2020)}]{browaeys2020many}%
  \BibitemOpen
  \bibfield  {author} {\bibinfo {author} {\bibfnamefont {A.}~\bibnamefont {Browaeys}}\ and\ \bibinfo {author} {\bibfnamefont {T.}~\bibnamefont {Lahaye}},\ }\bibfield  {title} {\bibinfo {title} {Many-body physics with individually controlled {R}ydberg atoms},\ }\href {https://www.nature.com/articles/s41567-019-0733-z} {\bibfield  {journal} {\bibinfo  {journal} {Nature Physics}\ }\textbf {\bibinfo {volume} {16}},\ \bibinfo {pages} {132} (\bibinfo {year} {2020})}\BibitemShut {NoStop}%
\bibitem [{\citenamefont {Dobrzyniecki}\ \emph {et~al.}(2024)\citenamefont {Dobrzyniecki}, \citenamefont {Heim},\ and\ \citenamefont {Tomza}}]{dobrzyniecki2024tunable}%
  \BibitemOpen
  \bibfield  {author} {\bibinfo {author} {\bibfnamefont {J.}~\bibnamefont {Dobrzyniecki}}, \bibinfo {author} {\bibfnamefont {P.}~\bibnamefont {Heim}},\ and\ \bibinfo {author} {\bibfnamefont {M.}~\bibnamefont {Tomza}},\ }\bibfield  {title} {\bibinfo {title} {Tunable two-species spin models with {R}ydberg atoms in circular and elliptical states},\ }\href {https://arxiv.org/abs/2411.14854} {\bibfield  {journal} {\bibinfo  {journal} {arXiv preprint arXiv:2411.14854}\ } (\bibinfo {year} {2024})}\BibitemShut {NoStop}%
\bibitem [{\citenamefont {Ding}\ \emph {et~al.}(2022)\citenamefont {Ding}, \citenamefont {Liu}, \citenamefont {Shi}, \citenamefont {Guo}, \citenamefont {M{\o}lmer},\ and\ \citenamefont {Adams}}]{Ding2022EnhancedMA}%
  \BibitemOpen
  \bibfield  {author} {\bibinfo {author} {\bibfnamefont {D.}~\bibnamefont {Ding}}, \bibinfo {author} {\bibfnamefont {Z.}~\bibnamefont {Liu}}, \bibinfo {author} {\bibfnamefont {B.}~\bibnamefont {Shi}}, \bibinfo {author} {\bibfnamefont {G.}~\bibnamefont {Guo}}, \bibinfo {author} {\bibfnamefont {K.}~\bibnamefont {M{\o}lmer}},\ and\ \bibinfo {author} {\bibfnamefont {C.~S.}\ \bibnamefont {Adams}},\ }\bibfield  {title} {\bibinfo {title} {Enhanced metrology at the critical point of a many-body {R}ydberg atomic system},\ }\href {https://api.semanticscholar.org/CorpusID:251040538} {\bibfield  {journal} {\bibinfo  {journal} {Nature Physics}\ }\textbf {\bibinfo {volume} {18}},\ \bibinfo {pages} {1447 } (\bibinfo {year} {2022})}\BibitemShut {NoStop}%
\end{thebibliography}%

\end{document}